\documentclass[12pt,a4paper]{article}
\usepackage{epsf}

\newcommand{\la}{\mathrel{\mathpalette\fun <}}

\def\fun#1#2{\lower3.6pt\vbox{\baselineskip0pt\lineskip.9pt
  \ialign{$\mathsurround=0pt#1\hfil##\hfil$\crcr#2\crcr\sim\crcr}}}
\newcommand{\dd}{\mathrm{d}}
\newcommand{\vecc}[1]{\mbox{\boldmath $#1$}}

\newcommand{\Tr}{\mathop\mathrm{Tr}}
\newcommand{\dirac}{\not\!}			    
\newcommand{\ddirac}{\not\!\!}			    

\title{Triplet Production by \\ Linearly Polarized Photons}

\author{I.V.~Akushevich$^{\,a}$, H. Anlauf$^{\,b}$, \'E.A. Kuraev$^{\,c}$,\\
  P.G. Ratcliffe$^{\,d}$ and B.G. Shaikhatdenov$^{\,c,}$\thanks{On leave
    of absence from the Institute of Physics and Technology, Almaty}
\\[4mm]
  \small\sl
  \begin{tabular}{l}
  $^a$ NC PHEP, Minsk, 220040, Belarus \\
  $^b$ Fachbereich Physik, Siegen University, 57068, Siegen, Germany \\
  $^c$ Joint Institute for Nuclear Research, 141980, Dubna, Russia \\
  $^d$ Dip.\ di Scienze, Universit\`a degli Studi dell'Insubria,
       sede di Como,\\
  \hphantom{$^a$}
       via Lucini 3, 22100 Como, Italy \\
  \hphantom{$^a$}
       and Istituto Nazionale di Fisica Nucleare, sezione di Milano
  \end{tabular}
}

\date{September 1999}

\begin{document}

\maketitle

\begin{abstract}
  The process of electron-positron pair production by linearly polarized
  photons is used as a polarimeter to perform mobile measurement of linear
  photon polarization. In the limit of high photon energies, $\omega$, the
  distributions of the recoil-electron momentum and azimuthal angle do not
  depend on the photon energy in the laboratory frame. We calculate the power
  corrections of order $m/\omega$ to the above distributions and estimate the
  deviation from the asymptotic result for various values of $\omega$.
\end{abstract}

\section{Introduction}

The differential cross-section for electron-positron pair production by
linearly polarized photons was derived in a series of papers during the period
1970--1972 \cite{Boldyshev1} (see also \cite{Vinokurov,Boldyshev2} and
references therein). Expressed as a function of $s=2m\omega$, where $m$ is the
electron mass (which we shall set equal to unity) and $\omega$ is the photon
energy in the laboratory reference frame, the differential cross-section with
respect to the azimuthal angle, $\phi$, between the photon polarization vector
$\mathrm{e}$ and the plane containing the initial-photon and recoil-electron
momenta, is given by
\begin{equation} \label{eq1}
  \frac{\dd\sigma}{\dd\phi}^\mathrm{asym}
  =
  \frac{\alpha^3}{m^2} \left[ \frac{28}{9}L
  - \frac{218}{27} - P \left( \frac{4}{9}L - \frac{20}{27} \right) \right],
\end{equation}
with
\[
  P = \xi_1\sin(2\phi) + \xi_3\cos(2\phi),
  \qquad
  L = \ln\frac{s}{m^2}.
\]
Here $\xi_1$ and $\xi_3$ are the Stokes parameters describing the photon
polarization, introduced through its spin-density matrix:
\[
  \rho_{ij}
  = \overline{{\mathrm e}_i{\mathrm e}_j}
  = \frac{1}{2} (1 + \vecc\sigma\vecc\xi)_{ij}.
\]

In the derivation of (\ref{eq1}) terms of order $m^2/s$ were systematically
neglected. The main contribution, $\sim\mathcal{O}(L)$, arises from
configurations with small recoil momentum,
$q=m\frac{2\cos\theta}{\sin^2\theta}\ll m$, where $\theta$ is the polar angle
of the recoil electron (i.e., the angle between the initial photon and
recoil-electron directions). However, the corresponding events presumably
cannot be measured experimentally. For the region $q\sim m$
($\theta\sim50^\circ$), the doubly differential cross-section was obtained in
\cite{Vinokurov}:
\begin{equation} \label{eq2}
  2\pi \frac{\dd^2\sigma}{\dd q\,\dd\phi}^\mathrm{\!\!\!asym}
  =
  \frac{2\alpha r_0^2}{3} \frac{q}{\varepsilon(\varepsilon-1)^2}
  \left[ a_0 - b_0 P \right],
\end{equation}
with
\[
  r_0 = \frac{\alpha}{m},
\qquad
  a_0 = 1 + \frac{2\varepsilon-3}{q} \ln(q+\varepsilon),
\qquad
  b_0 = 1 - \frac{1}{q} \ln(q+\varepsilon),
\]
where $\varepsilon=\sqrt{q^2+1}$. The comparatively large magnitude of the
azimuthal asymmetry
\begin{equation}
  \mathcal{A}
  = \frac{b_0}{a_0}
  = \frac{1}{7} - \frac{1}{245} q^2 + \frac{51}{34300} q^4 + O(q^6)
  \sim 14\%
\end{equation}
is, in fact, the reason this process is used for the polarimetry of linearly
polarized photons \cite{Boldyshev2,Work}.

The aim of the present paper is to calculate the power corrections of order
$1/s$ to the asymptotic expression for the asymmetry. The calculation of
radiative corrections to the asymptotic expression for the asymmetry is a
rather difficult problem, which we shall not touch here. A rough estimate gives
$\Delta\mathcal{A}^\mathrm{rad}\sim\frac{\alpha}{\pi}L\sim2-3\%$.

The differential cross-section of electron-positron pair photoproduction off a
free electron in the Born approximation is described by eight Feynman diagrams.
It was calculated numerically in particular by K.~Mork \cite{Mork}. The closed
expression for the unpolarized case is very cumbersome and was first obtained
in a complete form by E.~Haug during the period 1975--1985 \cite{Haug}. To the
best of our knowledge, the exact analytical expression for the differential
cross-section in the case of a polarized photon has not yet been published.
Special attention has been paid to the so-called Bethe-Heitler (BH) subset of
Feynman diagrams, whose contribution does not vanish in the high-energy limit,
$s\to\infty$ \cite{Suh}. The power corrections to this contribution to the
total cross-section, behaving as $L^3/s$ \cite{Borsellino}, indicate the need
for the exact expression.

A detailed analysis of the expressions of Haug's work reveals that the
interference terms of the BH matrix elements with the other three
gauge-invariant subsets (which take into account the bremsstrahlung mechanism
of pair creation and Fermi statistics for fermions) turn out to be of the order
of some percent for $s>50-60m^2$. On the other hand, the difference between the
asymptotic and the exact expression is still found to be of the order of
several percent for $s>3000m^2$, i.e., very far above threshold, rendering the
asymptotic expression useless in the energy range of interest. Arguments of
positivity of the cross-section provide the relevant upper bound for the
polarized part of the differential cross-section.

In Ref.~\cite{Endo} a Monte Carlo simulation of the process under consideration
was performed using the HELAS code, in which all eight lowest order diagrams
can be numerically treated without approximation. There it was shown that one
might consider only the two leading graphs in a wide range of photon energies
from 50 to 550~MeV. Note that this observation was made earlier for the
unpolarized case by Haug \cite{Haug} (who presented his results in explicit
analytical form).

Our paper is organized as follows. After introducing the problem, in
section~\ref{sec:kinematics} we analyze the kinematics of the process and give
a general expression for the differential cross-section taking into account
only leading and non-leading ($\sim1/s$) contributions. Section~\ref{sec:cross}
is devoted to the derivation of the differential cross-section with respect to
the azimuthal angle and the recoil momentum of the electron. In the concluding
section we present the correction to the cross-section and asymmetry, together
with some numerical estimates. Some details of the calculation may be found in
the Appendix.

\section{Kinematics and differential cross-section}
\label{sec:kinematics}

\begin{figure}[t]
\begin{tabular}{cccc}
\begin{picture}(80,100)
\put(39,75){\makebox(0,0){$K$}}
\put(39,15){\makebox(0,0){$p$}}
\put(90,15){\makebox(0,0){$q$}}
\put(65,65){\makebox(0,0){$-k_1$}}
\put(60,45){\makebox(0,0){$k$}}
\put(93,80){\makebox(0,0){$p_-$}}
\put(93,50){\makebox(0,0){$-p_+$}}
\put(0,0){
\epsfxsize=4cm
\epsfysize=4cm
\epsfbox{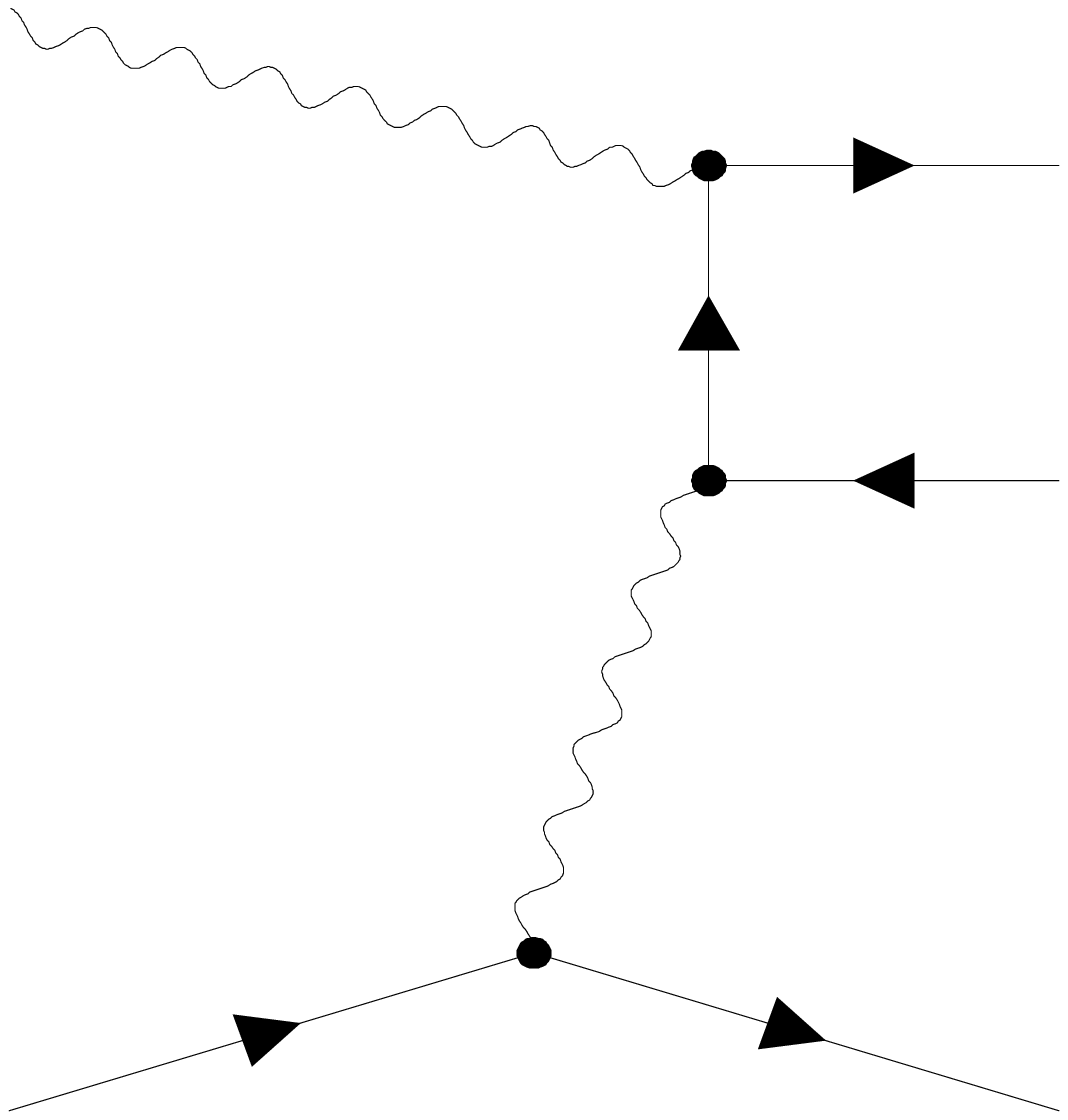}
}
\put(50,-10){\makebox(0,0){\small $M_1$}}
\end{picture}
&
\begin{picture}(80,100)
\put(39,72){\makebox(0,0){$K$}}
\put(39,15){\makebox(0,0){$p$}}
\put(90,15){\makebox(0,0){$q$}}
\put(93,80){\makebox(0,0){$p_-$}}
\put(93,50){\makebox(0,0){$-p_+$}}
\put(0,0){
\epsfxsize=4cm
\epsfysize=4cm
\epsfbox{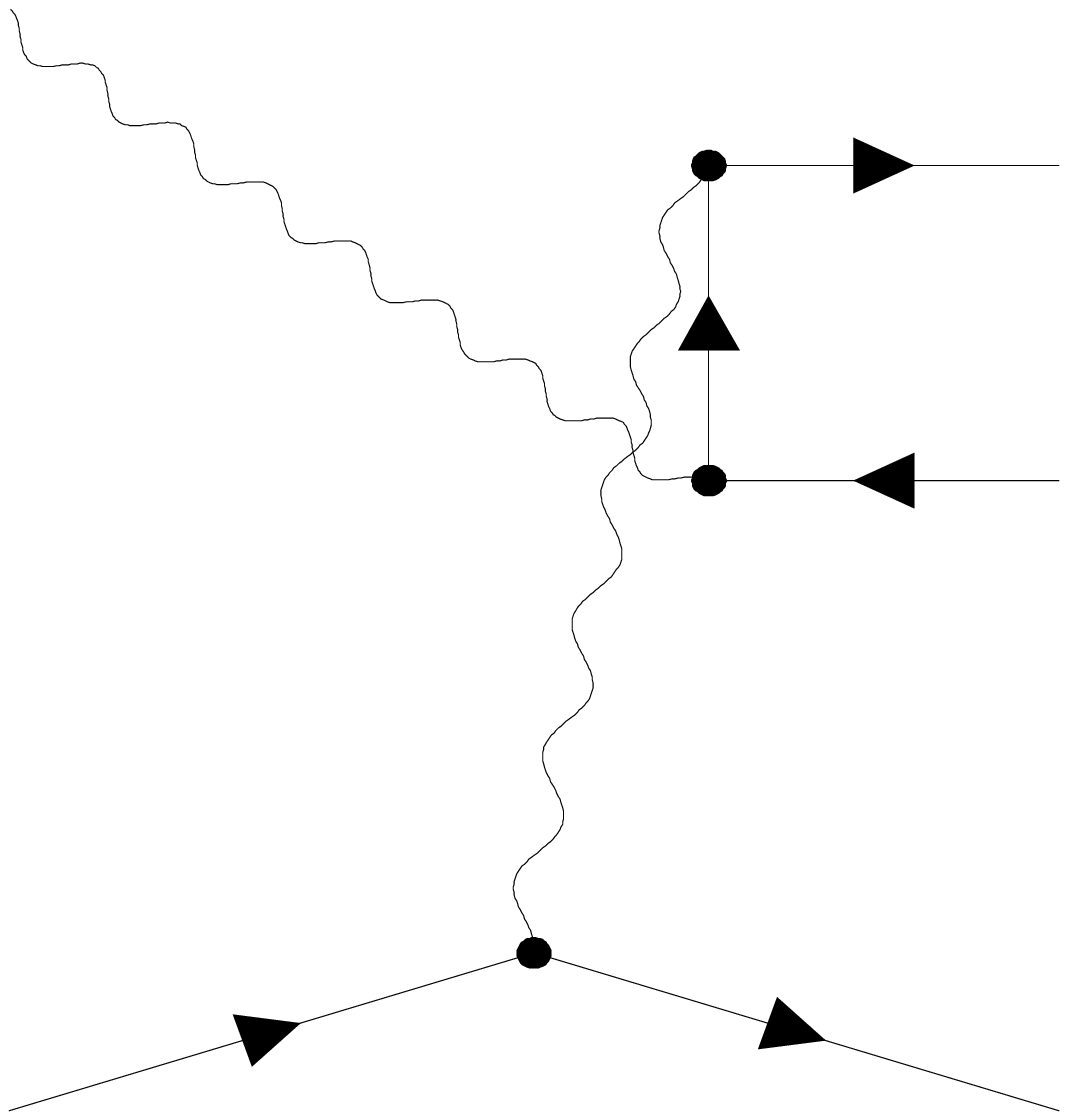}
}
\put(50,-10){\makebox(0,0){\small $M_2$}}
\end{picture}
&
\begin{picture}(80,100)
\put(39,75){\makebox(0,0){$K$}}
\put(39,15){\makebox(0,0){$p$}}
\put(90,15){\makebox(0,0){$p_-$}}
\put(93,80){\makebox(0,0){$q$}}
\put(93,50){\makebox(0,0){$-p_+$}}
\put(0,0){
\epsfxsize=4cm
\epsfysize=4cm
\epsfbox{graph1.eps}
}
\put(50,-10){\makebox(0,0){\small $M_3$}}
\end{picture}
&
\begin{picture}(80,100)
\put(39,72){\makebox(0,0){$K$}}
\put(39,15){\makebox(0,0){$p$}}
\put(90,15){\makebox(0,0){$p_-$}}
\put(93,80){\makebox(0,0){$q$}}
\put(93,50){\makebox(0,0){$-p_+$}}
\put(0,0){
\epsfxsize=4cm
\epsfysize=4cm
\epsfbox{graph2.eps}
}
\put(50,-10){\makebox(0,0){\small $M_4$}}
\end{picture}
\\[1cm]
\begin{picture}(80,100)
\put(45,75){\makebox(0,0){$K$}}
\put(39,15){\makebox(0,0){$p$}}
\put(90,15){\makebox(0,0){$q$}}
\put(93,80){\makebox(0,0){$p_-$}}
\put(93,50){\makebox(0,0){$-p_+$}}
\put(0,0){
\epsfxsize=4cm
\epsfysize=4cm
\epsfbox{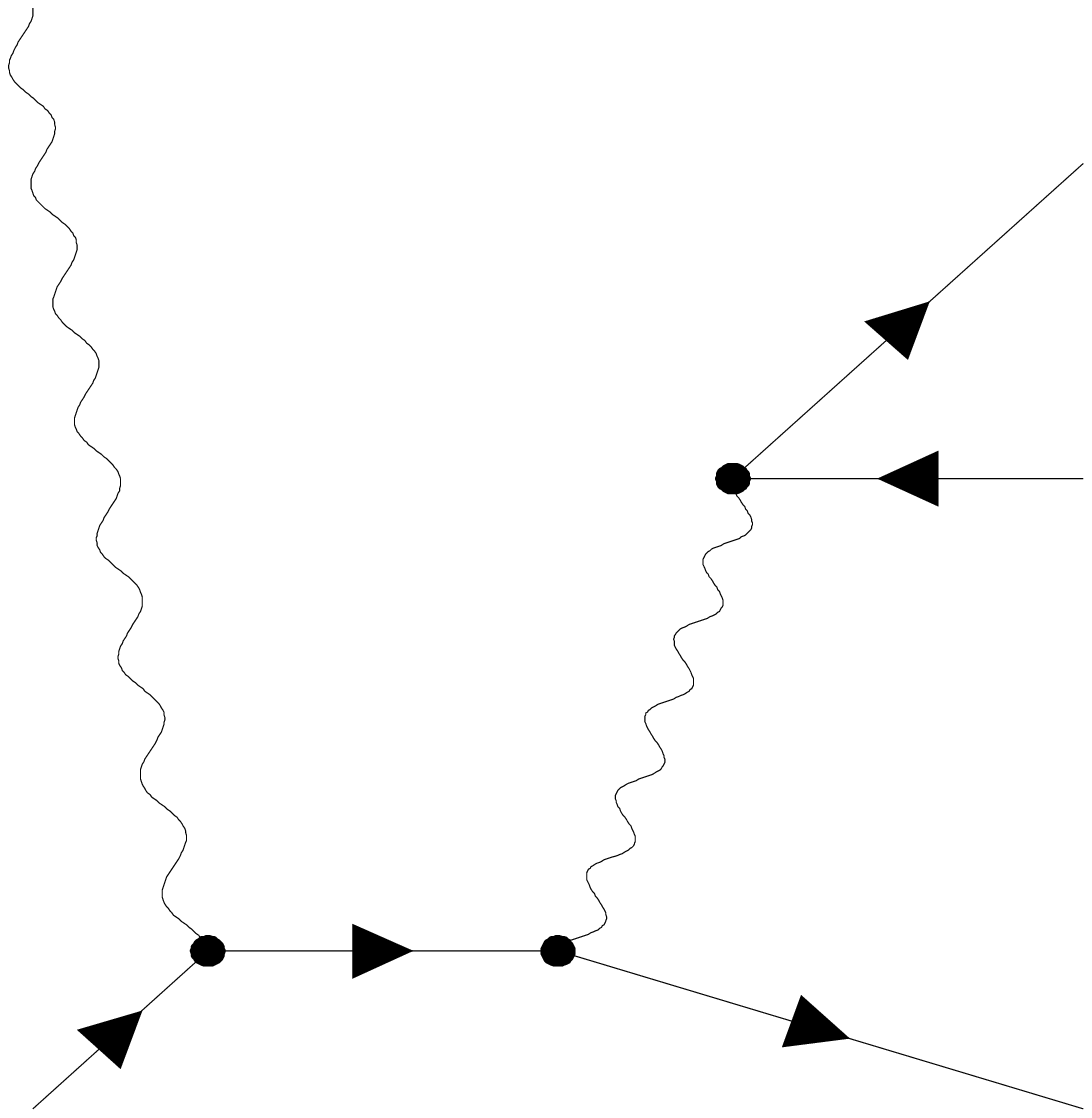}
}
\put(50,-10){\makebox(0,0){\small $M_5$}}
\end{picture}
&
\begin{picture}(80,100)
\put(47,77){\makebox(0,0){$K$}}
\put(39,15){\makebox(0,0){$p$}}
\put(90,15){\makebox(0,0){$q$}}
\put(93,80){\makebox(0,0){$p_-$}}
\put(93,50){\makebox(0,0){$-p_+$}}
\put(0,0){
\epsfxsize=4cm
\epsfysize=4cm
\epsfbox{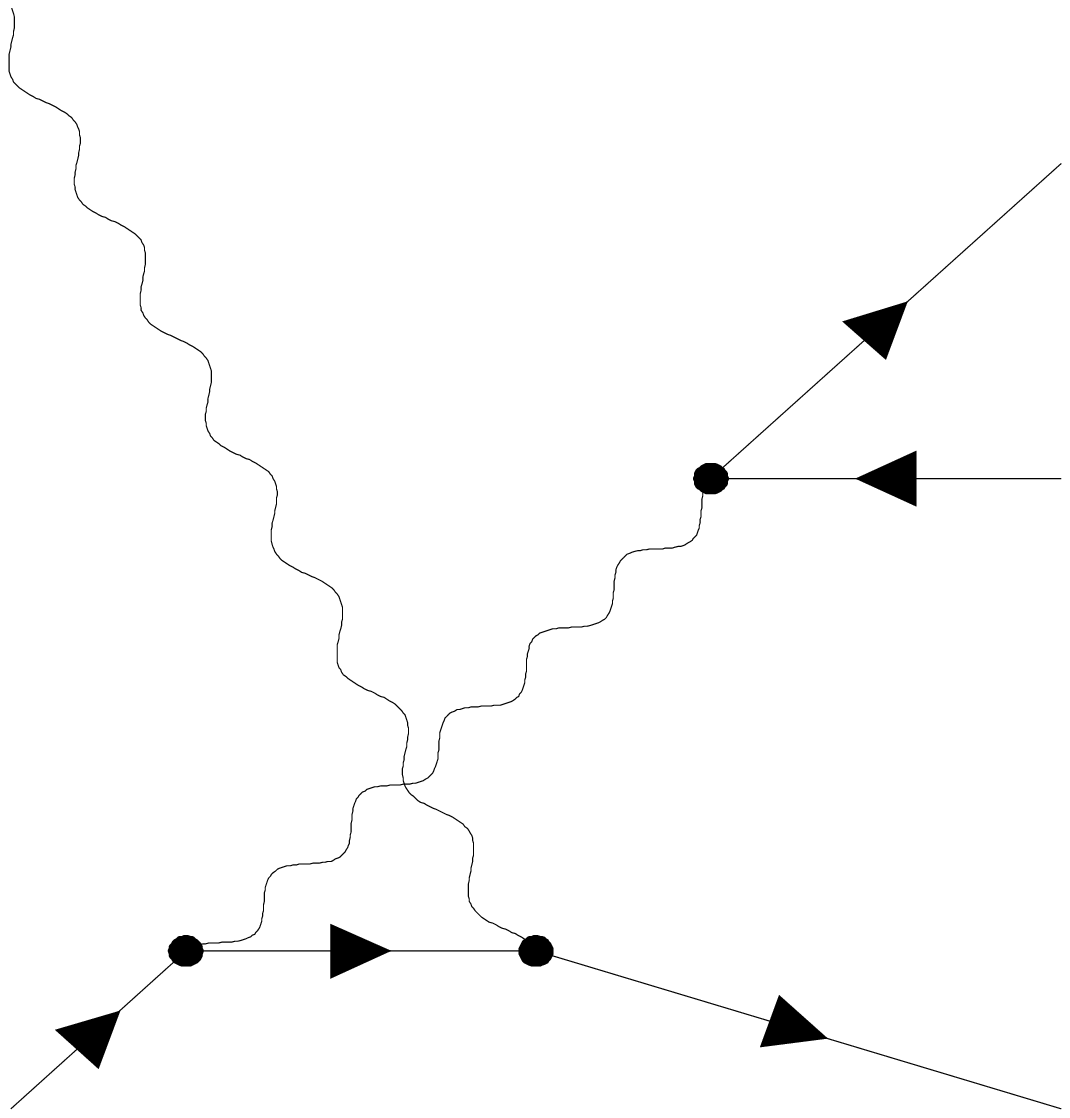}
}
\put(50,-10){\makebox(0,0){\small $M_6$}}
\end{picture}
&
\begin{picture}(80,100)
\put(45,75){\makebox(0,0){$K$}}
\put(39,15){\makebox(0,0){$p$}}
\put(90,15){\makebox(0,0){$p_-$}}
\put(93,80){\makebox(0,0){$q$}}
\put(93,50){\makebox(0,0){$-p_+$}}
\put(0,0){
\epsfxsize=4cm
\epsfysize=4cm
\epsfbox{graph3.eps}
}
\put(50,-10){\makebox(0,0){\small $M_7$}}
\end{picture}
&
\begin{picture}(80,100)
\put(47,77){\makebox(0,0){$K$}}
\put(39,15){\makebox(0,0){$p$}}
\put(90,15){\makebox(0,0){$p_-$}}
\put(93,80){\makebox(0,0){$q$}}
\put(93,50){\makebox(0,0){$-p_+$}}
\put(0,0){
\epsfxsize=4cm
\epsfysize=4cm
\epsfbox{graph4.eps}
}
\put(50,-10){\makebox(0,0){\small $M_8$}}
\end{picture}
\end{tabular}
\vspace{0.5cm}
\caption{The Feynman diagrams contributing to triplet production}
\label{Feyn}
\end{figure}

In the Born approximation, the cross-section for the process of pair production
off an electron,
\begin{equation}
  \gamma(K,\mathrm{e}) + e(p) \to e(q) + e(p_-) + e(p_+),
\end{equation}
with
\[
  q = K - k_1,
\qquad
  p_- = p - k,
\qquad
  p_+ = k_1 + k,
\]
is described by eight Feynman diagrams (Fig.~\ref{Feyn}), which can be combined
into four gauge-invariant subsets. Bearing in mind the desired application to
the case of high photon energies, $\omega\gg m$, we shall present the total
differential cross-section with leading terms (non-vanishing in the limit
$s=2m\omega\to\infty$) and terms of order $1/s$ (non-leading contributions).
The first arise from the BH subset, denoted by the indices (12), whereas the
non-leading terms come from interference of the BH amplitude with the sets
denoted (34), (56) and (78), as well as from the BH amplitude itself.

We use the following Sudakov decomposition of the momenta in our problem:
\begin{eqnarray}
  k = \alpha_k p' + \beta_k K + k_\bot,
  &&
  k_1 =	 \alpha_1 p' + x K + k_{1\bot},
\nonumber \\
  p' = p - K \frac{m^2}{s},
  &&
  s = 2pK,
\end{eqnarray}
with the properties that
\[
  p'^2 = K^2 = 0,
\qquad
  2p'p = 1,
\qquad
  k_\bot p = k_\bot K = 0.
\]
The energy-momentum fractions of the created pair are $x$ and $1-x$ while
$-\vec{k}_1$ and $\vec{k}-\vec{k}_1$ are their momentum components transverse
to the photon beam axis. The three mass-shell conditions,
\begin{eqnarray}
  (K-k_1)^2-1 = -s\alpha_1(1-x) - A = 0,
  &&
  A = \vec{k}_1^2 + 1,
\nonumber \\
  (k_1+k)^2-1 = s(\alpha_k + \alpha_1)(\beta_k + x) - B = 0,
  &&
  B = (\vec{k}+\vec{k}_1)^2 + 1,
\\
  (p-k)^2-1 = -s\beta_k(1 - \alpha_k) - \vec{k}^2 - \alpha_k = 0,
\nonumber
\end{eqnarray}
permit elimination of the following three Sudakov parameters:
\begin{equation}
  \beta_k = -\frac{\vec{k}^2}{s},
\qquad
  \alpha_k = \frac{1}{s} (s_1 + \vec{k}^2),
\qquad
  \alpha_1 = -\frac{A}{s(1-x)}.
\end{equation}
Henceforth, we neglect terms contributing to the cross-section at order
$\sim1/s^2$. Keeping in mind the application to experiment, we shall assume
\begin{equation} \label{eq8}
  \vec{k}^2 \sim 1.
\end{equation}
In this context, it should be noted that the above expressions for $\alpha_k$
and $\beta_k$ are also valid only to order $\mathcal{O}(s^{-2})$. Here
$s_1=(k+K)^2$ denotes the invariant mass squared of the created pair and
$\vec{k}^2$, the recoil-electron four-momentum transfer squared,
\begin{equation}
  s_1 = \frac{1}{(x+\beta_k)(1-x)}
  \left[ \left(\vec{k}_1 + (1-x) \vec{k} \right)^2 + 1 \right],
\qquad
  k^2 = -\frac{\vec{k}^2}{1-\alpha_k}.
\end{equation}

The recoil-electron 3-momentum, $q$, is related to its component transverse to
the photon beam axis, $|\vec{k}|$, as follows:
\begin{equation} \label{eq10}
  q^2 = \vec{k}^2 + \frac{1}{4} \vec{k}^4,
\quad
  \vec{k}^2 = q^2 \sin^2\theta = 2(\varepsilon - 1).
\end{equation}
The final-state phase volume may be expressed as
\begin{eqnarray}
  \dd\Gamma
  &=&
  \dd^4k \, \dd^4k_1 \;
  \delta((K-k_1)^2-1) \, \delta((k+k_1)^2-1) \, \delta((p-k)^2-1)
\nonumber \\
  &=&
  \frac{\dd^2k \, \dd^2k_1 \, \dd x}{4s(1-\alpha_k)(x+\beta_k)(1-x)}
\nonumber \\
  &=&
  \frac{\dd^2k \, \dd^2k_1 \, \dd x}{4sx(1-x)}
  \left( 1 + \alpha_k - \frac{\beta_k}{x} \right).
\end{eqnarray}
In terms of these variables, the total differential cross-section may be
written in the form:
\begin{eqnarray} \label{eq12}
  \dd\sigma
  &=&
  \frac{\alpha^3}{\pi^2(\vec{k}^2)^2}
  \biggl\{
  a^0_{1212} + \frac{1}{s}
  \biggl[
  a_{1212}^0 \biggl( -s\alpha_k - s\frac{\beta_k}{x} \biggr) + a_{1212}^1
\\
  && \qquad \mbox{}
  - \frac{2\vec{k}^2}{1-x}     a_{1234}
  + \frac{2\vec{k}^2}{x+\beta_k} a_{1278}
  - \frac{2\vec{k}^2}{s_1}     a_{1256}
  \biggr]
  \biggr\} \,
  \dd^2k_1 \, \dd x \, \dd^2k,
\nonumber
\end{eqnarray}
with
\begin{eqnarray} \label{eqai}
  a^0_{1212}
  &=&
  \frac{\vec{k}^2}{AB} - 4x(1-x) \frac{R_{11}(B-A)^2}{A^2B^2}
  + 8x(1-x) \frac{R_1(B-A)}{AB^2}
\nonumber \\
  && \quad \mbox{}
  - 4x(1-x) \frac{R}{B^2},
\nonumber \\
  a^1_{1212}
  &=& \mbox{}
  \frac{\vec{k}^2}{(x+\beta_k)}
  \biggl( \frac{B-A}{AB} - \frac{\vec{k}^2}{AB} \biggr)
\nonumber \\
  && \quad \mbox{}
  + 4\vec{k}^2
  \biggl(
  (3x-2) \frac{R_{11}(B-A)}{AB^2} + (1-x) \frac{R_{11}(B-A)}{A^2B}
  \biggr)
\nonumber \\
  && \quad \mbox{}
  - 4(3x-2) R \frac{\vec{k}^2}{B^2}
  - 4{\vec{k}^2}
  \biggl( (6x-4) \frac{R_1}{B^2} + (3-4x) \frac{R_1}{AB} \biggr),
\nonumber \\
  a_{1234}
  &=&
  \frac{1}{4} \biggl( \frac{B-A}{AB} - \frac{\vec{k}^2}{AB} \biggr)
  -2x(1-x) \frac{R_{11}(B-A)}{AB^2} + 2x(1-x) \frac{R}{B^2}
\nonumber \\
  && \quad \mbox{}
  -2x(1-x) \biggl( \frac{R_1(B-A)}{AB^2} - \frac{R_1}{B^2} \biggr),
\nonumber \\
  a_{1256}
  &=&
  \frac{1}{2(x+\beta_k)(1-x)}
  \biggl(
  x \frac{A}{B}
  + (1-2x)
  - x \frac{\vec{k}^2}{B}
  - (1-x) \biggl( \frac{B}{A} - \frac{\vec{k}^2}{A} \biggr)
  \biggr)
\nonumber \\
  && \quad \mbox{}
  + 4 \frac{R_{11}(B-A)}{AB}
  - 4 (1-x) \frac{R}{B}
  + 4 (1-x) \frac{R_1}{A}
  - 4 (2-x) \frac{R_1}{B},
\nonumber \\
  a_{1278}
  &=&
  - \frac{1}{4} \biggl( \frac{B-A}{AB} + \frac{\vec{k}^2}{AB} \biggr)
  + 2x(1-x) \frac{R_{11}(B-A)}{A^2B}
\nonumber \\
  && \quad \mbox{}
  - 2x(1-x) \frac{R_1}{AB}
\end{eqnarray}
and
\[
  R_{11} = \mathrm{e} k_1 \; \mathrm{e}^*k_1,
\quad
  R_{1} = \frac{1}{2}(\mathrm{e} k_1 \; \mathrm{e}^*k+\mathrm{e} k \;
  \mathrm{e}^*k_1),
\quad
  R = \mathrm{e}k \; \mathrm{e}^*k = \frac{\vec{k}^2}{2}(1+P).
\]
In general, the limits of variation for the parameters of the created pair are
imposed by experimental cuts together with the following relations
\begin{eqnarray*}
  (K-k_1)_0 = w(1-x) > m,
  &&
  (k_1+k)_0 = (x+\beta_k)\omega > m, \\
  s_1 < s,
  &&
  s = 2\omega
\end{eqnarray*}
or
\[
  \epsilon = \frac{2m^2}{s} < (x+\beta_k,1-x),
\qquad
  0 < \vec{k}_1^2 < s(x+\beta_k)(1-x) = \Lambda.
\]

\section{The inclusive distribution of the recoil electron}
\label{sec:cross}

The leading and non-leading contributions to the inclusive cross-section in
recoil-electron momentum may be organized as
\[
  2\pi \frac{\dd\sigma}{\dd q\,\dd\phi}
  =
  \frac{q}{\varepsilon(\varepsilon -1)^2} \alpha r_0^2 (I^\ell_{1212}+I^n),
  \qquad
  r_0 = \frac{\alpha}{m},
\]
where $I^\ell_{1212}$ comes from the first term in Eq.~(\ref{eq12}) and $I^n$
corresponds to the second. The leading and associated non-leading contributions
may be decomposed as
\begin{eqnarray}
  I^\ell_{1212}
  &=&
  \int\limits_{\epsilon}^{1-\epsilon}\dd x
  \int\limits_{\vec{k}_1^2<\Lambda} \frac{\dd^2k_1}{\pi} a^0_{1212}
\\
  &=&
  \int\limits_{\epsilon}^{1-\epsilon} \dd x
  \biggl\{ \int\limits_{0 < \vec{k}_1^2 < \infty}
	 - \int\limits_{\Lambda < \vec{k}_1^2 < \infty}
  \biggr\}
  \frac{\dd^2k_1}{\pi} a^0_{1212}
  =
  I^{\ell_1}_{1212} + I^{\ell_2}_{1212}.
\nonumber
\end{eqnarray}
Using the table of integrals provided in the Appendix, for the first term in
brackets we obtain
\begin{eqnarray} \label{eq13}
  I^{\ell_1}_{1212}
  =
  \int\limits_0^1 \dd z \int\limits_{\epsilon-\beta_k}^{1-\epsilon} \dd x 
  \biggl\{
  \frac{\vec{k}^2}{\gamma} + 8x(1-x)
  \biggl[ 1 - \frac{4+\vec{k}^2}{4\gamma} - R \frac{z(1-z)}{\gamma} \biggr]
  \biggr\}
\nonumber \\
  = \int\limits_0^1 \dd z
  \biggl\{
  \frac{\vec{k}^2}{\gamma} (1-2\epsilon+\beta_k) 
  + \frac{4}{3}
  \biggl[ 1 - \frac{4+\vec{k}^2}{4\gamma} - R \frac{z(1-z)}{\gamma} \biggr]
  \biggr\},
\end{eqnarray}
where $\gamma=1+z(1-z)\vec{k}^2$. For $\epsilon=0$, we reproduce the result
given in Eq.~(\ref{eq2}). In order to see this, one may use the expansion
\[
  \int\limits_0^1 \frac{\dd z}{\gamma}
  =
  1 - \frac{1}{6} \vec{k}^2 + \frac{1}{30} \vec{k}^4 + \ldots
\]
and the relation between $\vec{k}^2$ and $q^2$ given above in Eq.~(\ref{eq10}).
The second term, $I^{\ell_2}_{1212}$, may be calculated using the expansion of
$a^0_{1212}$ for $\vec{k}_1^2\gg1$ (see Eq.~(\ref{eq29}) in the Appendix)
\begin{eqnarray} \label{eq16}
  I^{\ell_2}_{1212} = - \frac{2\vec{k}^2}{s} (L - \ln 2 - 1).
\end{eqnarray}

The quantity $I^n$ may also be expressed as a sum:
$I^n=I^c_{1212}+I^\mathrm{int}$. Consider now the contributions arising from
corrections to the leading term (see Eq.~(\ref{eq12})):
\begin{eqnarray} \label{eq17}
  I_{1212}^c
  &=&
  \int\limits_{\epsilon}^{1-\epsilon} \dd x
  \int\limits_{\vec{k}_1^2<\Lambda} \frac{\dd^2k_1}{\pi} \;
  a^0_{1212} \biggl( - \alpha_k - \frac{\beta_k}{x} \biggr)
\nonumber \\
  &=&
  \int\limits_{\epsilon}^{1-\epsilon} \dd x
  \int \frac{\dd^2k_1}{\pi} \;
  a^0_{1212} \frac{(1-x)\vec{k}^2}{sx}
\nonumber \\
  && \mbox{}
  - \frac{1}{s}
  \int\limits_{\epsilon}^{1-\epsilon} \frac{\dd x}{x(1-x)}
  \int\limits_{\vec{k}_1^2<\Lambda} \frac{\dd^2k_1}{\pi} \;
  a^0_{1212} \left[ 1 + \left( \vec{k}_1 + (1-x)\vec{k} \right)^2 \right]
\nonumber \\
  &=&
  I^{c_1}_{1212} + I^{c_2}_{1212} + I^{c_3}_{1212}.
\end{eqnarray}
The first term on the RHS of Eq.~(\ref{eq17}) gives
\[
  I^{c_1}_{1212} = \frac{\vec{k}^2}{s}			 
  \int\limits_0^1{\dd z}
  \biggl\{
  \frac{\vec{k}^2}{\gamma} (L - \ln2 - 1)
  + \frac{8}{3}
  \biggl[ 1 - \frac{4+\vec{k}^2}{4\gamma} - R \frac{z(1-z)}{\gamma} \biggr]
  \biggr\}.
\]

It is convenient also to present the second term as a sum of two parts. The
first, containing $L^2$, comes from the $a^0_{1212}$ term, non-vanishing for
both $x\to0$ and $x\to1$:
\begin{eqnarray}
  I^{c_2}_{1212}
  &=&
  -\frac{\vec{k}^2}{s}
  \int\limits_{\epsilon}^{1-\epsilon} \frac{\dd x}{x(1-x)}
  \int \frac{\dd^2k_1}{\pi} \frac{xA+(1-x)B-x(1-x)\vec{k}^2}{AB}
\nonumber \\
  &=&
  \frac{\vec{k}^2}{s} \int\limits_0^1{\dd z} \frac{\vec{k}^2}{\gamma}
  - 2 \frac{\vec{k}^2}{s}
  \biggl[ \frac{1}{2} (L^2 - \ln^22) - \frac{\pi^2}{6} \biggr].
\end{eqnarray}
The remaining terms are
\begin{eqnarray*}
  I^{c_3}_{1212}
  &=&
  -\frac{4}{s}	       
  \int\limits_{\epsilon}^{1-\epsilon} \dd x
  \int_{\vec{k}_1^2<\Lambda}
  \frac{\dd^2k_1}{\pi} \,
  \biggl[ xA + (1-x)B - x(1-x)\vec{k}^2 \biggr]
  \nonumber \\
  && \qquad \qquad
  \biggl[
  - R_{11} \biggl( \frac{1}{A} - \frac{1}{B} \biggr)^2
  + 2R_1 \biggl( \frac{1}{AB} - \frac{1}{B^2} \biggr) - \frac{R}{B^2}
  \biggr]
  =
  r_1 + r_2.
\end{eqnarray*}
The last term in the first brackets, which gives rise to $r_2$, is ultraviolet
convergent upon integration over $\vec{k}_1$,
\begin{eqnarray}
  r_2 = -\frac{4\vec{k}^2}{3s} \int\limits_0^1 \dd z
  \biggl[ -1 + \frac{4+\vec{k}^2}{4\gamma} + \frac{z(1-z)}{\gamma} R \biggr],
\end{eqnarray}
whereas the other is
\begin{eqnarray}
  r_1
  =
  \frac{2}{s} \int\limits_0^1 \dd z
  \biggl\{ \biggl( L - \frac{7}{2} \biggr) \vec{k}^2 + R \biggr\}.   
\end{eqnarray}
The non-leading term, $a^1_{1212}$, along with the interference contribution
(see Eq.~(\ref{eq12})) to the inclusive cross-section yields
\begin{eqnarray} \label{eq15}
  I^\mathrm{int}
  =
  \frac{\vec{k}^2}{s} \int\limits_0^1 \dd z
  \biggl\{
  - \frac{\vec{k}^2}{\gamma} (L-\ln 2) - 4
  + \frac{4+\vec{k}^2}{\gamma} + \frac{4z(1-z)}{\gamma} R
  \biggr\}.
\end{eqnarray}
The contribution of the structure $a_{1256}$ is exactly zero to order $1/s$. To
make this clear, one can bring it into the following form,
\begin{eqnarray*}
  a_{1256}
  &=&
  \frac{1}{2x(1-x)A} \biggl\{ (1-x) (\vec{k}^2+A-B)
\\
  && \qquad \qquad \quad \; \mbox{}
  + 8x(1-x) (R_{11}+(1-x)R_1) \biggr\}
  - \Biggl(
  \begin{array}{c}
    x \leftrightarrow 1-x \\
    \vec{k}_1 \to \vec{k}_1+\vec{k}
  \end{array}
  \Biggr),
\end{eqnarray*}
which is antisymmetric with respect to interchange of $x$ and $1-x$.

\section{Conclusions}

Our result for the order $1/s$ correction to the inclusive cross-section is the
sum of the expressions for $I^{\ell_1,\ell_2}_{1212}$, $I^{c_1,c_2,c_3}_{1212}$
and $I^n$ given in Eqs.~(\ref{eq13}--\ref{eq15}). Organizing the inclusive
cross-section in the form
\[
  2\pi \frac{\dd \sigma}{\dd q\,\dd\phi}
  =
  \frac{2q}{3\varepsilon(\varepsilon-1)^2} \alpha r_0^2
  \biggl[
  a_0 + \frac{2(\varepsilon-1)}{s} a_1
  - P \biggl( b_0 + \frac{2(\varepsilon-1)}{s} b_1 \biggr)
  \biggr],
\]
where $a_0$ and $b_0$ are given above in Eq.~(\ref{eq2}) and
\begin{eqnarray*}
  a_1 = \frac{3}{2} \biggl[ -L^2 + C - 2L_q (\varepsilon+1) \biggr],  
  &&
  b_1 = -\frac{3}{2},						      
\\
  C = \ln^22 + 2\ln2 + \frac{\pi^2}{3} - 4 \approx 1.1566,	      
  &&
  L_q = \frac{\ln(\varepsilon+q)}{q},				      
\end{eqnarray*}
we extract the asymmetry,
\begin{eqnarray}
  \mathcal{A}
  = \mathcal{A} + \Delta\mathcal{A}
  = \frac{b_0}{a_0} + \frac{\vec{k}^2}{s} \frac{b_1a_0-a_1b_0}{a_0^2}.
\end{eqnarray}

\begin{figure}[t]
\vspace{5cm}
\begin{picture}(150,180)
\put(30,0){
\epsfxsize=12cm
\epsfysize=16cm
\epsfbox{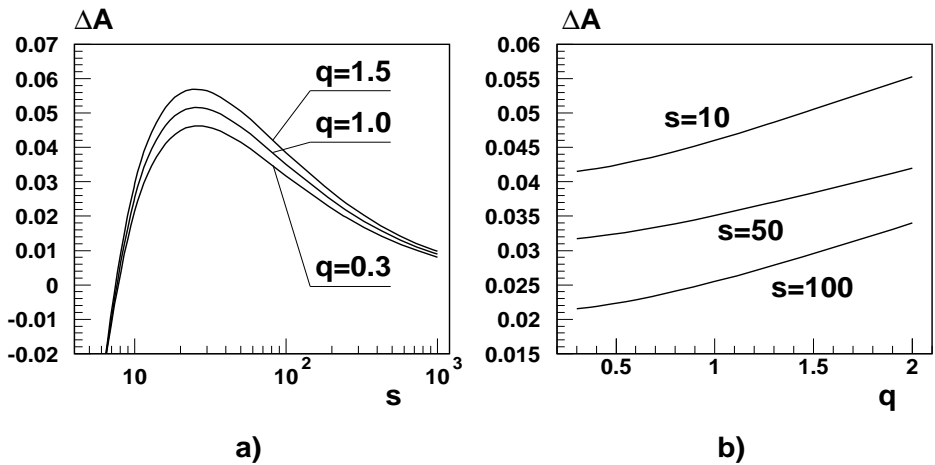}
}
\end{picture}
\vspace{-7.5cm}
\caption{The dependence of the correction, $\Delta\mathcal{A}$, to the
  asymmetry as a function of a) $s$ and b) $q$.}
\label{fig2}
\end{figure}

The expansion of $\Delta\mathcal{A}$ can be recast in the form (and this is our
final result),
\begin{eqnarray}
  \Delta\mathcal{A} = \frac{3(\varepsilon-1)}{sa_0^2}	      
  \biggl[ L^2 - C - 1 + L_q (5-L^2+C) - 2L_q^2 (\varepsilon+1) \biggr].
\end{eqnarray}
The the dependence of this quantity on $q$ at fixed $s$ and vice versa is shown
in Figs.~\ref{fig2}a and b (recall that we have set $m=1$).

For small enough $s\la10$ the terms of order of $1/s^n$, for $n\ge2$, become
essential and the approach presented in this paper is not applicable.

We should like to point out that our results are in qualitative agreement with
those obtained using the HELAS code \cite{Endo} and reported in the talk
delivered at the workshop \cite{Work}.

\section*{Acknowledgments}
Three of us (IVA, EAK and BGS) are grateful to the DESY staff for hospitality.
The work of EAK and BGS was partially supported by the Heisenberg-Landau
Programme and the Russian Foundation for Basic Research grant 99-02-17730. The
work of HA was supported by the Bundesministerium f\"ur Bildung, Wissenschaft,
Forschung und Technologie (BMBF), Germany. EAK is also grateful to
L.S.~Petrusha for help.

\section*{Appendix}

Since the amplitudes of the gauge-invariant sets of diagrams (3,4), (5,6) and
(7,8) are suppressed by at least one power of $\vec{k}^2/s$ as compared to
amplitude (1,2), we need only consider interference terms. Thus, for the
modulus of the matrix element, squared and summed over fermion spin states, we
have within leading ($\sim s^2$) and non-leading ($\sim s$) accuracy, in order
\begin{eqnarray}
  \sum|M|^2
  =
  \frac{(4\pi\alpha)^3}{\vec{k}^2}
  \left\{
    - \frac{2T_{1234}}{s(1-x)}
    - \frac{2T_{1256}}{s_1}
    + \frac{2T_{1278}}{sx}
    + \frac{T_{1212}(1-\alpha_k)^2}{\vec{k}^2}
  \right\},
\end{eqnarray}
where
\begin{eqnarray}
  T_{1212}
  &=&
  \Tr \left\{ (\dirac q+1) \gamma_\mu (\dirac p+1) \gamma_{\nu} \right\}
\nonumber \\
  &&
  \times \Tr \left\{
  (\ddirac K-\dirac k_1+1) O_{12}^{\mu\lambda} (\dirac k_1+\dirac k-1)
  \overline{O}_{34}^{\nu\sigma} \right\}
  e^{\vphantom{*}}_\lambda e^*_\sigma,
\nonumber \\
  T_{1234}
  &=&
  \Tr \left\{ (\dirac q+1) \gamma_\mu (\dirac p+1) \gamma_{\nu}
  (\ddirac K-\dirac k_1+1) O_{12}^{\mu\lambda} (\dirac k_1+\dirac k-1)
  \overline{O}_{34}^{\nu\sigma} \right\}
  e^{\vphantom{*}}_\lambda e^*_\sigma,
\nonumber \\
  T_{1256}
  &=&
  \Tr \left\{ (\dirac q+1)\gamma_\mu (\dirac p+1)
  \overline{O}_{56}^{\nu\sigma} \right\} \\ \nonumber
  &&
  \times \Tr \left\{ (\ddirac K-\dirac k_1+1) O_{12}^{\mu\lambda}
  (\dirac k_1+\dirac k-1) \gamma_\nu \right\}
  e^{\vphantom{*}}_\lambda e^*_\sigma,
\\ \nonumber
  T_{1278}
  &=&
  \Tr \left\{ (\dirac q+1) \gamma_\mu (\dirac p+1)
  \overline{O}_{78}^{\nu\sigma} (\ddirac K-\dirac k_1+1)
  O_{12}^{\mu\lambda} (\dirac k_1+\dirac k-1) \gamma_{\nu} \right\}
  e^{\vphantom{*}}_\lambda e^*_\sigma,
\\[7pt]
  O_{12}^{\mu\lambda}
  &=&
  - \frac{x+\beta_k}{B} \gamma_\mu (\ddirac K-\dirac k_1-\dirac k+1)
\gamma_\lambda
  - \frac{1-x}{A} \gamma_\lambda (-\dirac k_1+1) \gamma_\mu,
\nonumber \\
  O_{34}^{\mu\lambda}
  &=&
  -\frac{x}{B} \gamma_\mu (\ddirac K-\dirac k_1-\dirac k+1) \gamma_\lambda
  -\frac{1}{s} \gamma_\lambda (\dirac p-\ddirac K-\dirac k+1) \gamma_\mu,
\nonumber \\
  O_{56}^{\mu\lambda}
  &=&
  \frac{1}{s} \left[ -\gamma_\lambda (\dirac p-\ddirac K-\dirac k+1) \gamma_\mu
  + \gamma_\mu(\dirac p+\ddirac K+1) \gamma_\lambda \right],
\nonumber \\
  O_{78}^{\mu\lambda}
  &=&
  - \frac{1-x}{A} \gamma_\lambda (-\dirac k_1+1) \gamma_\mu
  + \frac{1}{s} \gamma_\mu (\dirac p+\ddirac K+1) \gamma_\lambda,
\nonumber
\end{eqnarray}
and $q=p-k$.

The quantities $T_{ijkl}$ are related to the $a_{ijkl}$ given in
Eq.~(\ref{eqai}) by
\begin{eqnarray}
  T_{1212}	&=& 16 s   x(1-x) \, [a_{1212}^0s + a_{1212}^1],
\nonumber \\
  T_{1234,1278} &=& 16 s^2 x(1-x) \, a_{1234,1278},
\nonumber \\
  T_{1256}	&=& 16 s   x(1-x) \, a_{1256}.
\end{eqnarray}
To perform the integration over $\vec{k}_1$, we introduce an ultraviolet
cut-off $\vec{k}_1^2<\Lambda$, which may be omitted when calculating convergent
integrals for the corrections. The integrals containing $A$ or $B$ are
(hereinafter we omit the terms of order of $1/s$)
\begin{eqnarray}
  \int \frac{\dd^2\vec{k}_1}{\pi} \frac{1}{A^2} =
  \int \frac{\dd^2\vec{k}_1}{\pi} \frac{1}{B^2} = 1,
  &&
  \int \frac{\dd^2\vec{k}_1}{\pi} \frac{R_1}{B^2} = -R,
\\
  \int \frac{\dd^2\vec{k}_1}{\pi} \frac{R_{11}}{A^2}
  =
  \frac{1}{2} (\ln\Lambda - 1),
  &&
  \int \frac{\dd^2\vec{k}_1}{\pi} \frac{R_{11}}{B^2}
  =
  \frac{1}{2} (\ln\Lambda - 1) + R.
\nonumber
\end{eqnarray}
In order to avoid linearly divergent integrals, we combine denominators as
follows:
\[
  \frac{1}{AB}
  =
  \int\limits_0^1 \frac{\dd z}{[(\vec{k}_1+z\vec{k})^2+\gamma]^2},
\]
with $\gamma=1+\vec{k}^2z(1-z)$. Integrating over $\vec{k}_1$ we obtain
\begin{eqnarray}
  \int \frac{\dd^2\vec{k}_1}{\pi} \frac{1}{AB}
  &=&
  \int\limits_0^1 \frac{\dd z}{\gamma}
  \;=\;
  \frac{1}{q} \ln(\varepsilon+q)
  \;=\;
  L_q,
\nonumber \\
  \int \frac{\dd^2\vec{k}_1}{\pi} \frac{R_1}{AB}
  &=&
  - \int\limits_0^1 \frac{z\dd z}{\gamma} R
  \;=\;
  - R \int\limits_0^1 \frac{\dd z}{2\gamma},
\\ \nonumber
  \int \frac{\dd^2\vec{k}_1}{\pi} \frac{R_{11}}{AB}
  &=&
  \int\limits_0^1 \dd z
  \biggl[
  \frac{1}{2} (\ln\Lambda-1) - \frac{1}{2} \ln\gamma + \frac{z^2}{\gamma} R
  \biggr].
\end{eqnarray}
To evaluate the quantity $r_1$, we use the following set of integrals:
\begin{eqnarray}
  \int \frac{\dd^2\vec{k}_1}{\pi}
  [xA + (1-x)B]\frac{R}{B^2}
  \hspace*{6.5ex}
  &=&
  R \, \left( \ln\Lambda + x\vec{k}^2 \right),
\nonumber \\
  \int \frac{\dd^2\vec{k}_1}{\pi}
  [xA + (1-x)B] R_1 \frac{A-B}{AB^2}
  \hspace*{0.5ex}
  &=&
  -R \, \left( \ln\Lambda - 1 + x\vec{k}^2 \right),
\\
  \int \frac{\dd^2\vec{k}_1}{\pi}
  [xA + (1-x)B] R_{11} \frac{A-B}{AB}
  &=&
  \left[ R + \frac{\vec{k}^2}{2} \right]
  \left[ \ln\Lambda - \frac{3}{2} \right]
  + x \vec{k}^2 R.
\nonumber
\end{eqnarray}
In deriving (\ref{eq16}), for $a^0_{1212}$ averaged over angles with
$\vec{k}_1^2\gg1$, we make use of
\begin{equation} \label{eq29}
  \overline{a^0_{1212}}|_{\vec{k}_1^2\gg 1}
  \approx
  \frac{\vec{k}^2}{(\vec{k}_1^2)^2} \left( 1  - 2x(1-x) \right)
\end{equation}
and to perform the angular averaging in $r_1$ we use
\[
  \overline{R_{11}(\vec{k}\vec{k}_1)^2}
  \to
  \frac{1}{8} (\vec{k}_1^2)^2 [\vec{k}^2 + 2R].
\]


\end{document}